\documentclass[preprintnumbers,prb,amsmath,amssymb]{revtex4}
\usepackage{graphicx}
\usepackage{bm}

\begin{document}

\title{
Approach towards quasi-monoenergetic laser ion acceleration with doped target
}

\author{Toshimasa Morita}

\affiliation{Quantum Beam Science Directorate, Japan Atomic Energy Agency,
8-1-7 Umemidai, Kizugawa, Kyoto 619-0215, Japan}


\begin{abstract}
Ion acceleration by using a laser pulse irradiating a disk target
which includes hydrogen and carbon
is examined using three-dimensional particle-in-cell simulations.
It is shown that over $200$ MeV protons can be generated
by using a $620$TW, $5\times10^{21}$ W/cm$^2$ laser pulse.
In a polyethylene (CH$_2$) target,
protons and carbon ions separate and form two layers
by radiation pressure acceleration.
A strong Coulomb explosion in this situation
and Coulomb repulsion between each layer generates high energy protons.
A doped target, low density hydrogen within a carbon disk,
becomes a double layer target which is comprised of a thin and low density
hydrogen disk on the surface of a high-$Z$ atom layer.
This then generates a quasi-monoenergetic proton beam.
\end{abstract}



\maketitle

\section{INTRODUCTION}

It was shown
that charged particles can be accelerated using an intense laser pulse
theoretically and via numerical simulations in 1979. \cite{TaDa}
The acceleration of electrons by laser light was proposed.
Ions, which are much heavier than electrons,
also can be accelerated by the laser irradiation.
However, the acceleration mechanism is different
from that of electron acceleration.
Studies on ion acceleration by a laser pulse started a little later
than the studies of the acceleration of electrons,
and has been active since around 2000.
It was reported in 2000 that with an
irradiated laser intensity of
$10^{19}$W/cm$^{2}$ onto a metallic thin film target
fast protons of about $20$MeV were observed, \cite{CLK} and
also in 2000 with an irradiated laser intensity of
$10^{20}$W/cm$^{2}$ onto a CH polymer target
fast protons of about $60$MeV were obtained. \cite{SNV}
Current important topics in the study of ion acceleration by a laser pulse are
the increase in ion energy and number of high energy ions,
and the production of a quasi-monoenergetic ion beam.
There are many applications using laser ion acceleration
which require high energy and quasi-monoenergetic ion beam. 
\cite{ROT,ESI1,BEE}
The achieved proton energy at present is not high for
some applications such as hadron therapy, which requires $200$MeV protons.
Therefore,
it is important to study conditions for generating higher energy ions
and higher quality, quasi-monoenergetic, ion beams with lower power
and energy lasers by using some special techniques.
\cite{BWP,DL,HAC,HSM,MEBKY,PPM,PRK,SPJ,Toncian,YAH,MEBKY2,TM}

In this paper,
I use three-dimensional (3D) particle-in-cell (PIC) simulations
to investigate how high energy, and high quality proton beams
can be generated by a several-hundred-terawatt laser.
Our aim is to obtain a high energy ($\mathcal{E} \geq 200$ MeV) and
high quality ($\Delta \mathcal{E}/\mathcal{E}_\mathrm{ave} \leq$ a few \%)
proton beam using the actual capabilities of current laser systems.
Therefore, I use a laser with a
peak power of $620$TW, energy of $18$J, and
peak intensity of $5\times 10^{21}$ W/cm$^{2}$ in the simulations.
I show a way to obtain $200$ MeV protons and quasi-monoenergetic ion beams
by using a target which consists of hydrogen and carbon
mixed uniformly.

As suggested in Ref. \cite{TM},
a target containing much hydrogen should be generate high energy protons.
Therefore, I take a polyethylene (CH$_2$) target
to generate high energy protons.
This is because polyethylene includes much hydrogen
and is easy to handle, since it is a solid at room temperature.
In Sec. \ref{poly},
it is shown that high energy protons are obtained by using this target.

In sec. \ref{dope}, I use a doped target (C$_n$H$_m$, $n \gg m$)
that has very low density hydrogen within a carbon disk.
This type of target can be produced to use controlled hydrogen doping of
a high-$Z$ material.
The ratio of the hydrogen in the target is much smaller than
in the previous case,
it is shown that we can obtain a quasi-monoenergetic proton beam.

\section{High energy proton beam} \label{poly}

In this section,
I show simulation results which obtain high energy protons,
$\mathcal{E} \approx 200$ MeV, by using a polyethylene (CH$_2$) target.

\subsection{Simulation model}

Here, the parameters used in the simulations are shown.
The spatial coordinates are normalized by the laser wavelength
$\lambda=0.8$ $\mu$m and
time is measured in terms of the laser period $T_0=\lambda/c$,
where $c$ is the speed of light.
I use an idealized model, in which a Gaussian linear polarized laser pulse
is incident on a disk target represented by a collisionless plasma.

The disk target has a diameter of $8\lambda $ and a thickness of $0.5\lambda $.
The electron density is $n_{e}=9\times 10^{22}$ cm$^{-3}$.
Therefore, the proton density is $2\times 10^{22}$ cm$^{-3}$,
the carbon ion density is $1\times 10^{22}$ cm$^{-3}$.
The total number of quasiparticles is $3\times 10^{8}$.
The number of grid cells is equal to $5000\times3000\times 3000$
along the $X$, $Y$, and $Z$ axes, respectively.
Correspondingly, the simulation box size is
$113\lambda \times 68\lambda \times 68\lambda$.
The boundary conditions for the particles and for the fields are
periodic in the transverse ($Y$,$Z$) directions and absorbing at the
boundaries of the computation box along the $X$ axis.
$xyz$ coordinates are used in the text and figures;
the origin of the coordinate system is located at the center of
the rear surface of the initial target,
and the directions of the $x$, $y$, and $z$ axes are the same as
those of the $X$, $Y$, and $Z$ axes, respectively.
That is, the $x$ axis denotes the direction perpendicular to the target surface
and the $y$ and $z$ axes lie in the plane of the target surface.
The ionization state of carbon ion is assumed to be $Z_{i}=+6$.

The laser pulse with dimensionless amplitude
$a_0=q_eE_{0}/m_{e}\omega c=50$,
which corresponds to a laser peak intensity of $5\times 10^{21}$ W/cm$^{2}$,
is $10\lambda $ long in the propagation direction, $27$ fs in duration,
and focused to a spot with size $4\lambda $ (FWHM), which corresponds
to a laser peak power of $620$ TW and a laser energy of $18$ J.

\subsection{Simulation results of CH$_2$ target}

\begin{figure}[tbp]
\includegraphics[clip,width=11.0cm,bb=53 5 522 212]{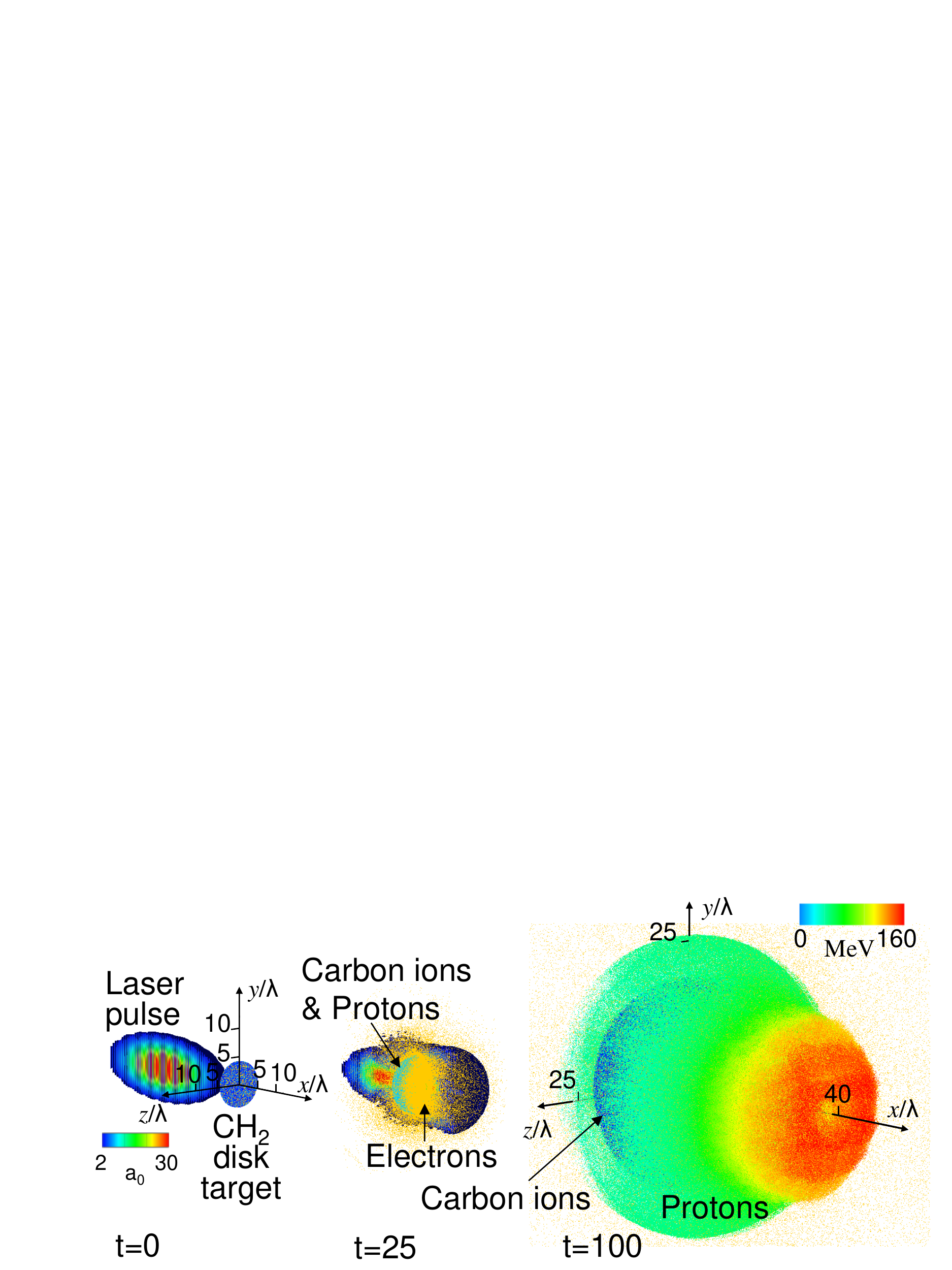}
\caption{
The laser pulse is normally incident on a polyethylene (CH$_2$) disk target.
3D view of the particle distribution and electric field magnitude
(isosurface for value $a=2$).
Half of the electric field box has been removed to reveal the internal
structure. For protons, the color corresponds to energy.
}
\label{fig:fig_n1}
\end{figure}

I show simulation results to obtain high energy proton beam
by using a polyethylene (CH$_2$) target.
Figure \ref{fig:fig_n1} shows the particle distribution and
the electric field magnitude in time with normal incidence.
The electric field is oriented in the $y$ direction.
It shows the initial shape of the target, $t=0$, and the laser pulse
and the interaction of the target and laser pulse, $t=25,100 T_0$.
Protons are classified by color in terms of energy.
At $t=25 T_0$, the laser pulse is just around the target and
has strong interactions with the target.
The target maintains its initial disk shape at this time.
After $t=25 T_0$,
the laser pulse passes through or reflects off of the target,
the proton and carbon ion cloud explode by Coulomb explosions
and grow in time.
The high energy protons are distributed at the
extreme of the ion cloud having an ellipsoid shape, $t=100 T_0$.

\begin{figure}[tbp]
\includegraphics[clip,width=9.0cm,bb=10 0 274 187]{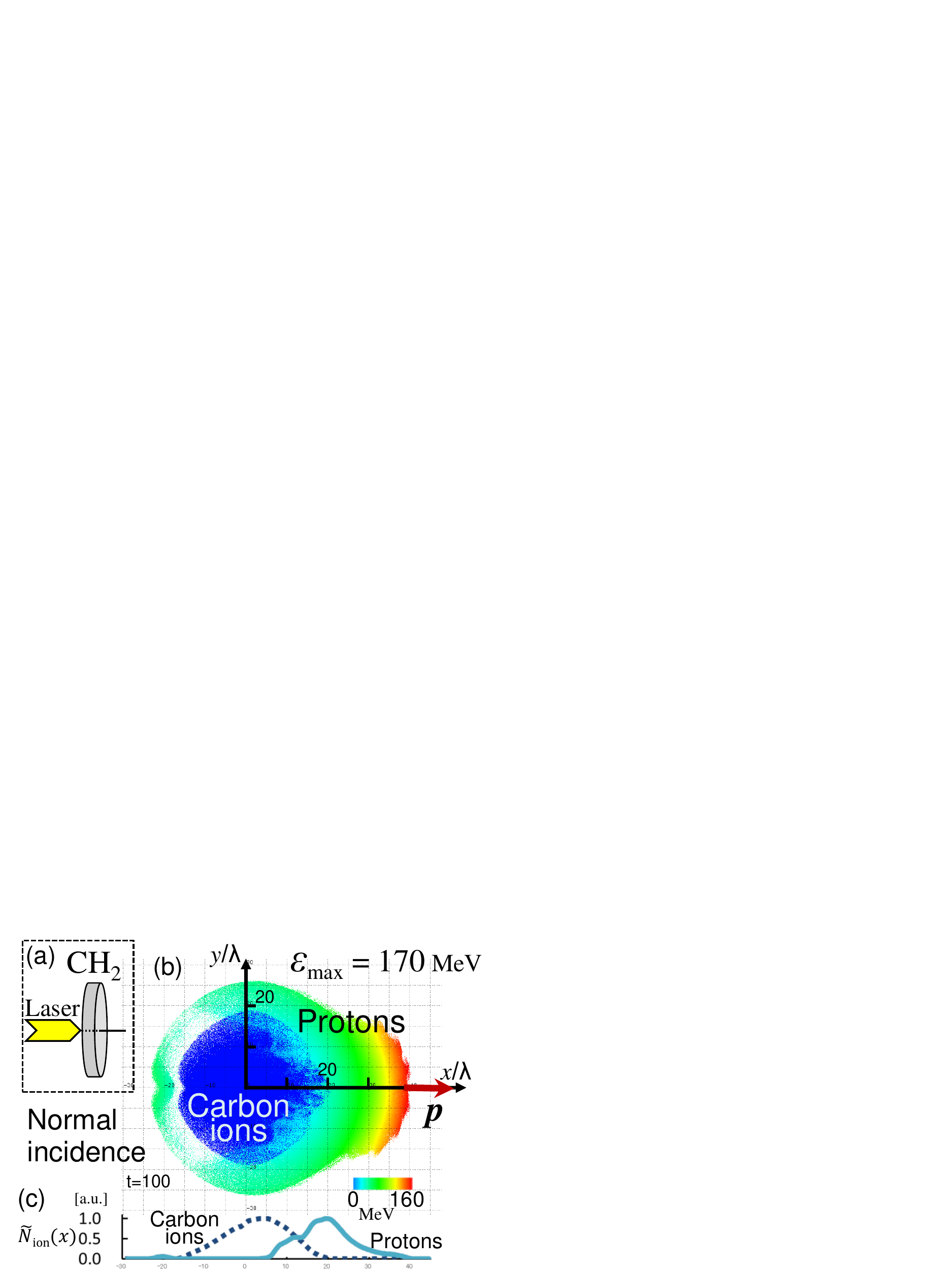}
\caption{
(a) The laser pulse is normally incident on a polyethylene disk target.
(b) Distribution of the carbon ions (dark color in the center)
and protons (color scale) at $t=100 T_0$.
Half of the ion cloud has been removed to reveal the internal structure;
a two-dimensional projection is shown looking along the $z$ axis.
The thick arrow shows the momentum vector, $\bm{p}$, of the high energy protons.
(c) Distribution of the number of carbon ions and protons in the $x$ direction.
$\tilde{N}_\mathrm{ion}$ is the number of ions per unit length along $x$,
and is normalized by its maximum value.
}
\label{fig:fig_n2}
\end{figure}

A cross section of the ion cloud in the $(x,y)$ plane at $t=100 T_0$
is shown in Fig. \ref{fig:fig_n2}(b).
Protons are classified by color in terms of energy.
Carbon ions are shown by dark color.
The proton and carbon ion cloud explode by Coulomb explosions.
The carbon ions are distributed around the center
and the protons are on the outer edge of it.
Moreover,
almost all the protons are distributed on $+x$ side of the carbon ions,
as seen in Fig. \ref{fig:fig_n2}(c).
On the other hand, the movement of the carbon ion cloud is small.
The $x$ coordinate of the center point of
the proton cloud is $18\lambda$ and the carbon ion cloud is $3\lambda$.
The vertical axis of Fig. \ref{fig:fig_n2}(c) is the number of ions
normalized by its maximum value,
$\tilde{N}_\mathrm{ion}(x)=N_\mathrm{ion}(x)/\max\{N_\mathrm{ion}(x)\}$.
The number of ions
$N_\mathrm{ion}(x)=
\int_{-\infty}^{\infty} \int_{-\infty}^{\infty} \rho(x,y,z)dydz$,
where $\rho(x,y,z)$ is the ion density.
The carbon ions and protons are clearly separated into different areas,
due to the different ``mass''.
I use the term ``mass'' to mean $m/q$ where $m$ is the mass of an ion
and $q$ is its charge.
Small-``mass'' ions will be called ``light'' and
big-``mass'' ions will be called ``heavy''.  \cite{TM}
The proton cloud has large movement in the $+x$ direction
by radiation pressure acceleration (RPA),
since a proton is ``lighter'' than a carbon ion.
In this layer separated situation,
the proton cloud receives charge repulsion from the carbon ion cloud in the
$+x$ direction and it efficiently accelerates the protons.
The high energy protons are achieved by RPA,
strong Coulomb explosion of both the ion cloud,
and charge repulsion from the carbon ion cloud.
The maximum proton energy, $\mathcal{E}_\mathrm{max}$, is 170 MeV.
Protons are distributed in different areas based on each energy level.
High-energy protons are distributed on the $+x$ side edge of the proton cloud
and are moving in the $+x$ direction.
The high energy proton momentum is almost along the $x$ axes.
The angle of the momentum vector of these protons, $\bm{p}$,
with the $x$ axes is $\phi \approx 0^\circ$.
I take the average of the highest $1\times 10^7$ protons
in determining this momentum vector.
These protons have a maximum energy of 174 MeV, minimum energy of 170 MeV.

\begin{figure}[tbp]
\includegraphics[clip,width=8.0cm,bb=9 4 530 405]{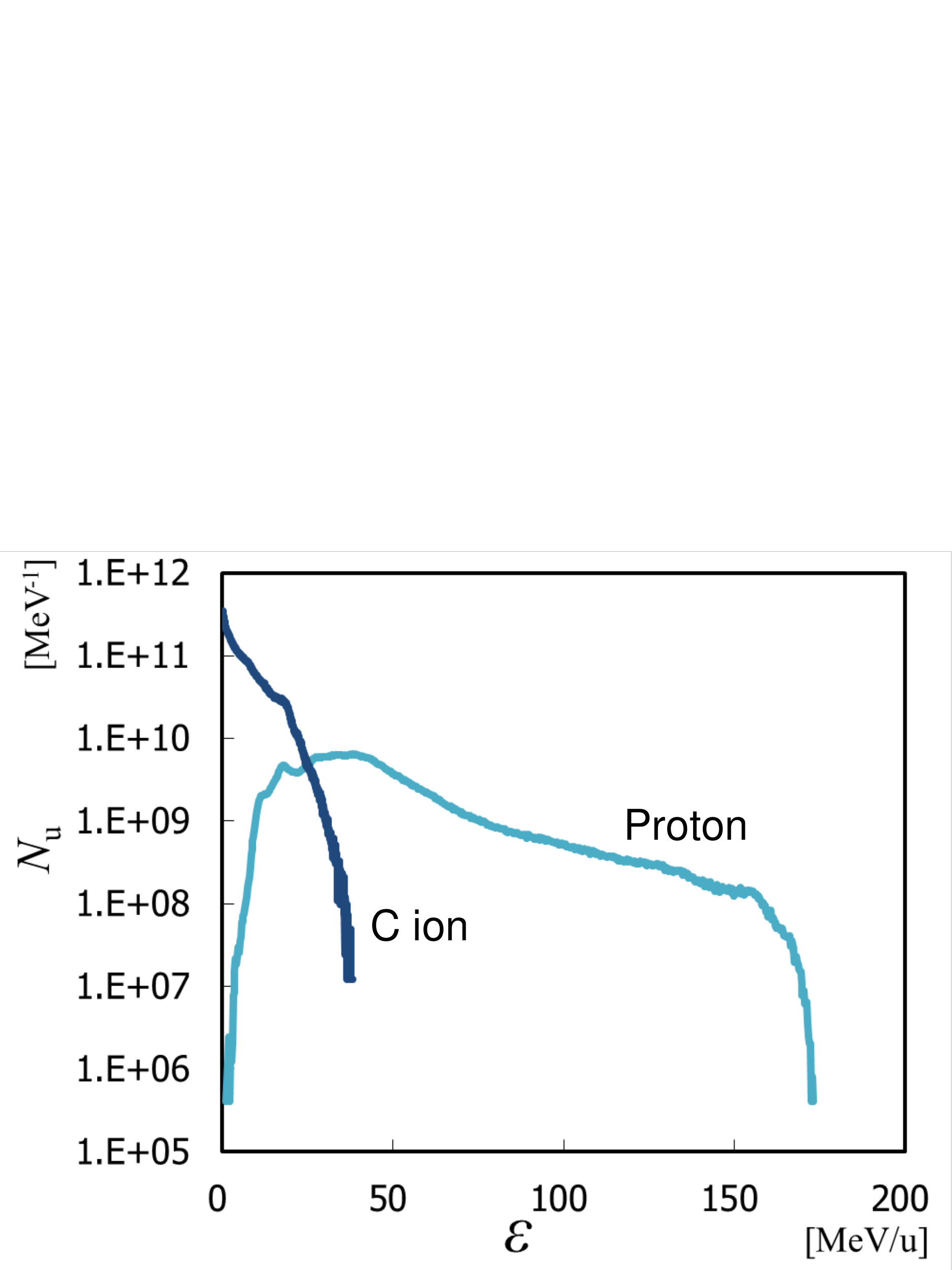}
\caption{
Energy spectrum of protons and carbon ions obtained in the simulation
at $t=100 T_0$.
}
\label{fig:fig_n3}
\end{figure}

Figure \ref{fig:fig_n3} shows the energy spectrum of the protons and carbon ions
for the normal incidence case at $t=100 T_0$.
The vertical axis is given in units of the number of nucleons, protons,
per $1$ MeV width.
The horizontal axis is the energy of a nucleon, proton, $\mathcal{E}$.
The number of protons is enough for some application (e.g.; hadron therapy).
We can select only high energy protons by cutting out its distributed area
with a shutter, or a magnet and a slit.
Then we obtain a high quality, a quasi-monoenegeric, proton beam. \cite{TM}
Since polyethylene has carbon and hydrogen,
we can also obtain
the accelerated carbon ions in addition to high energy protons.
However, the maximum energy of the carbon ions is $38$MeV/u
which is much lower than the proton one.
The energy spectrum of the carbon ions has a peak around $0$ MeV,
and proton one has a peak around $40$MeV.
This means, many carbon ions are almost not accelerated,
and are distributed around the initial target position.
On the other hand, most protons are efficiently accelerated.
Using the polyethylene target,
we will first observe high energy protons, then after of them,
at some later time, we will observe decent high energy carbon ions.

\begin{figure}[tbp]
\includegraphics[clip,width=8.0cm,bb=8 5 527 455]{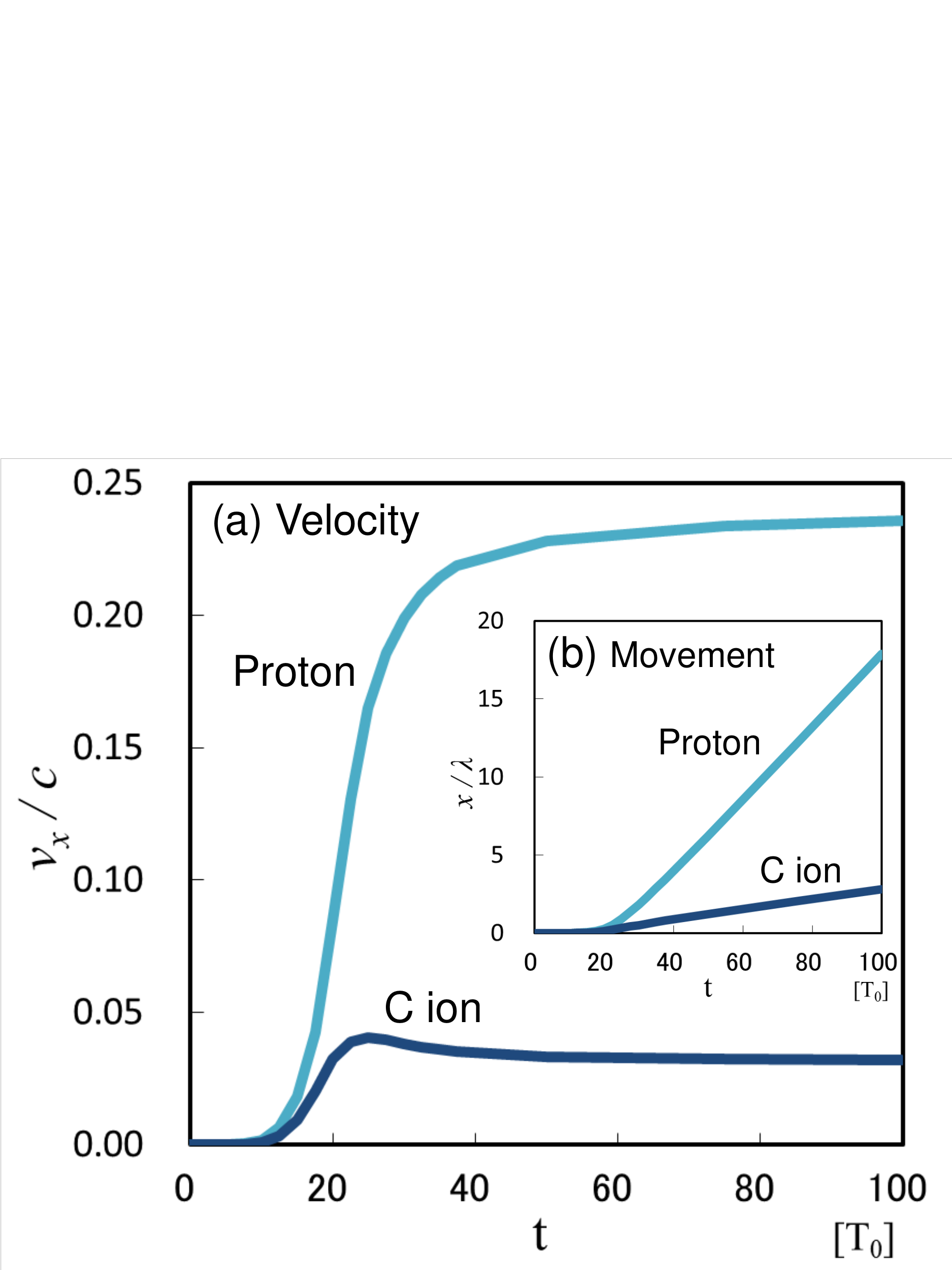}
\caption{
Velocity of the protons and carbon ions in the $x$ direction normalized
by the speed of light, $v_x/c$, as a function of time (a), and
movement of the protons and carbon ions in the $x$ direction normalized
by the wavelength, $x/\lambda$, (b).
}
\label{fig:fig_n4}
\end{figure}

I showed that the energy and distribution of each ion species
at $t=100T_0$ above.
Next,
is shown the variations in time and results at the early simulation times.
Figure \ref{fig:fig_n4} shows the velocity of the proton cloud and
the carbon ion cloud normalized by the speed of light and
the position of each ion cloud normalized by the wavelength,
in the $x$ direction as a function of time.
These are averaged values for all protons and all carbon ions.
The proton cloud, and even the carbon ion cloud,
get velocity in the $+x$ direction by RPA.
The proton cloud velocity is much higher than the carbon ion cloud one.
The proton cloud velocity rises rapidly at the initial time, $t\sim 30T_0$,
when the laser pulse is still around the target (see Fig. \ref{fig:fig_n1})
and the increasing of the velocity slows down at times $t > 30 T_0$,
after the laser pulse passes through or reflects off the target.
The velocity of the carbon ions has a peak of $v_x/c=0.04$ at $t=25T_0$
and decreases to $v_x/c=0.03$ around $t=50T_0$.
This is because, there is a proton cloud which has plus charge
in the front of the carbon ions.
Therefore, the ion cloud receives a negative, $-x$ direction,
force from the proton cloud, and decreases the velocity.
In other words, the momentum of the carbon cloud is transferred to the
proton cloud and protons get a higher energy.

\begin{figure}[tbp]
\includegraphics[clip,width=8.0cm,bb=5 0 540 642]{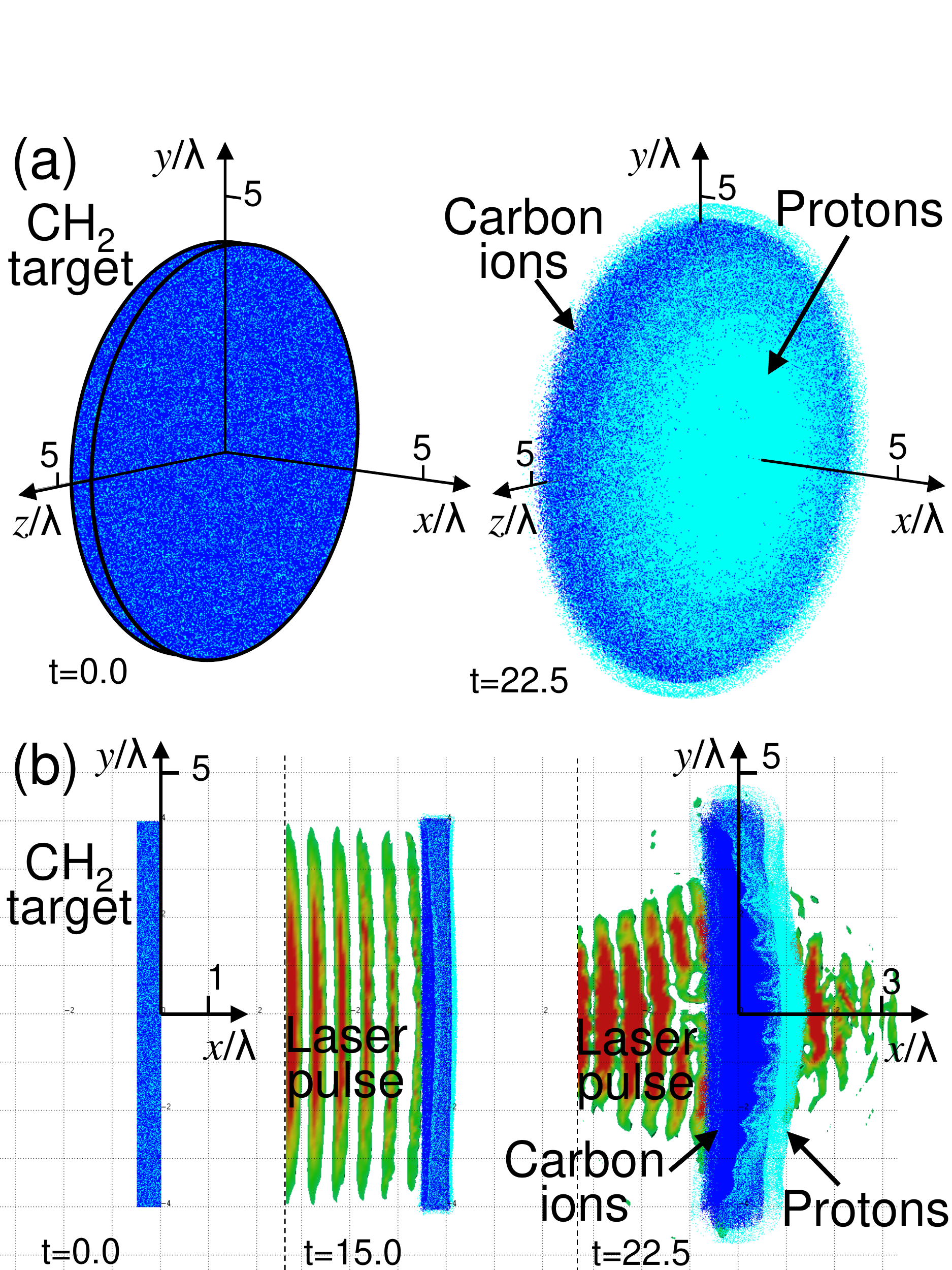}
\caption{
The initial target state ($t=0.0$),
and at an early simulation time ($t\leq 22.5T_0$).
(a) 3D view.
(b) 2D projection is shown looking along $z$ axis,
half of the box has been removed to reveal the internal structure;
electric field magnitude is also shown (isosurface for value $a_0=15$).
The proton layer position on the surface of the carbon ion layer
is shown at $t=22.5T_0$.
}
\label{fig:fig_n5}
\end{figure}

Figure \ref{fig:fig_n5} shows the target state at an early simulation time
$t < 25T_0$.
At this time, the laser pulse is still around the target.
It is shown that
the proton layer has moved in the $+x$ direction by RPA,
and the carbon ions almost stay at their initial position.
The amount of movement of the proton cloud is almost half of
the carbon layer thickness at $t=15.0T_0$.
Then, the carbon ions and protons are clearly separated into two layers
at $t=22.5T_0$.
The proton layer is placed on the surface of the carbon ion layer,
especially where the area around the center of the target,
with keep its initial thickness.
The proton cloud and the carbon ion cloud have the same sign of charge.
Therefore, in this separation situation into two layers,
the protons receive a force in the $+x$ direction from the carbon ion cloud
and protons are further accelerated by this force. 
By contrast, the carbon ions decrease their energy.
We can confirm this from the velocity decrease of the carbon ions around
$t=25T_0$ (see Fig. \ref{fig:fig_n4}).
If there was not this layer separation,
protons and carbon ions would mix uniformly,
this acceleration mechanism would not work,
and the obtained proton energy would decrease.
If the position of the proton layer and carbon ion layer was reverse,
it would have formed two layers in which the carbon layer was positioned
on the rear side surface of the proton layer.
This acceleration mechanism works negatively for the protons.
The obtained proton energy would decrease,
because the momentum of the proton cloud would transfer to the carbon ion cloud.
Therefore,
it is very important to generate high energy protons to form the
two layer target in which
the proton layer position is on the rear side surface of the carbon ion layer.
The RPA at the initial acceleration stage plays an important role to generate
high energy protons.
It is
a target which consists of mixed ``heavy'' atoms and ``light'' atoms
formulating the two layer target.
This is because ``light'' ions get a higher velocity than ``heavy'' ions
in the laser propagation direction by RPA.
In this effect, ``light'' ions and ``heavy'' ions separate and
form the two layer target in the initial stage of acceleration.

\begin{figure}[tbp]
\includegraphics[clip,width=9.0cm,bb=10 0 305 190]{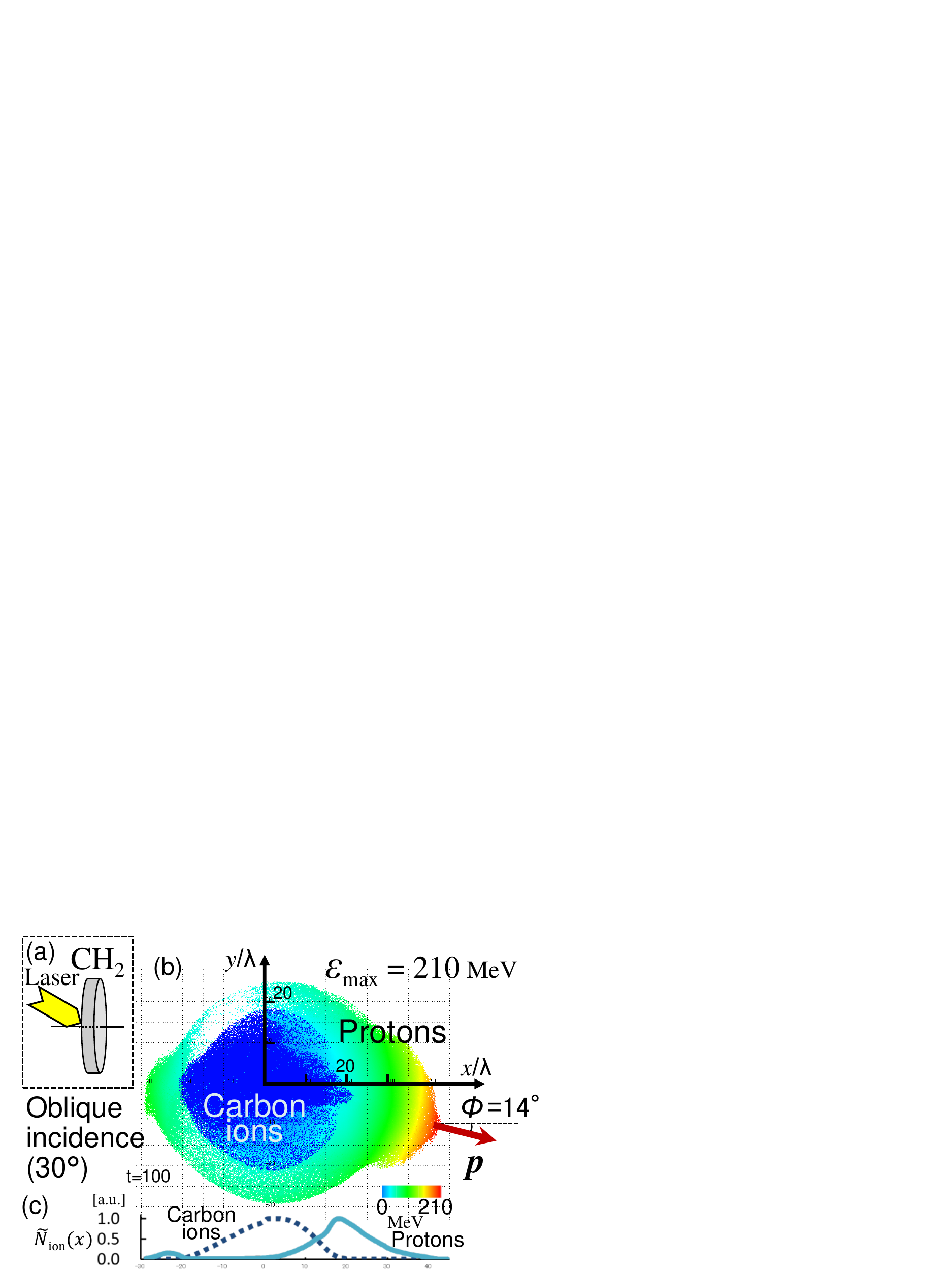}
\caption{
(a) The laser pulse is obliquely incident on a polyethylene disk target.
(b) Distribution of the carbon ions (dark color in the center)
and protons (color scale) at $t=100 T_0$.
Half of the ion cloud has been removed to reveal the internal structure;
a two-dimensional projection is shown looking along the $z$ axis.
The thick arrow shows the momentum vector, $\bm{p}$, of the high energy protons.
(c) Distribution of the number of carbon ions and protons in the $x$ direction.
$\tilde{N}_\mathrm{ion}$ is the number of ions per unit length along $x$,
and is normalized by its maximum value.
}
\label{fig:fig_o1}
\end{figure}

It was shown that
we can obtain higher energy protons by oblique incidence. \cite{MEBKY}
I show the result of the simulations of oblique incidence where
the incidence angle is $30^\circ$.
In the oblique incidence, a $p-$polarized laser pulse is incident on a target.
The other simulation parameters without the incidence angle
are same as the previous normal incidence case.
Figure \ref{fig:fig_o1} shows the same type of figure as Fig.\ref{fig:fig_n2},
the particle distribution at $t=100 T_0$.
The proton and carbon ion cloud are exploded by Coulomb explosion,
and these ion clouds move in the $+x$ direction, also.
However, high-energy protons are distributed at a place
which is shifted from the $x$ axes,
in the laser propagation direction. \cite{MEBKY,MEBKY2}
The maximum proton energy, $\mathcal{E}_\mathrm{max}$, is 210 MeV.
The deflection angle of the high energy protons is $\phi=14^\circ$.
I take the average of the highest $1\times 10^7$ protons
to determine this momentum vector.
These protons have a maximum energy of 218 MeV and minimum energy of 210 MeV.
We can obtain higher energy protons,  $\mathcal{E} >$ 200 MeV,
for oblique incidence.
This is $40$MeV higher than normal incidence.
Also the maximum energy of the carbon ions in the oblique incidence is
$42$MeV/u,
which is higher than normal incidence by $4$MeV/u.

\section{Quasi-monoenergetic proton beam} \label{dope}

I showed a target which consisted of initially uniformly mixed
carbon and hydrogen, and formed into a
two layer target by RPA in the previous section.
As suggested in Ref. \onlinecite{DL},
a quasi-monoenegeric proton beam can be obtained using targets
with a very thin and low density hydrogen layer on the surface of
a high-$Z$ atom layer (double layer target).
Therefore, if the hydrogen density in the target is much lower than high-$Z$
ion density, it should compose a double layer target
which has a thin and low density proton layer on the high-$Z$ ion layer.
Then we should obtain a quasi-monoenegeric proton beam.
Here, I show the simulation results from this point of view.

\begin{figure}[tbp]
\includegraphics[clip,width=10.0cm,bb=0 0 520 294]{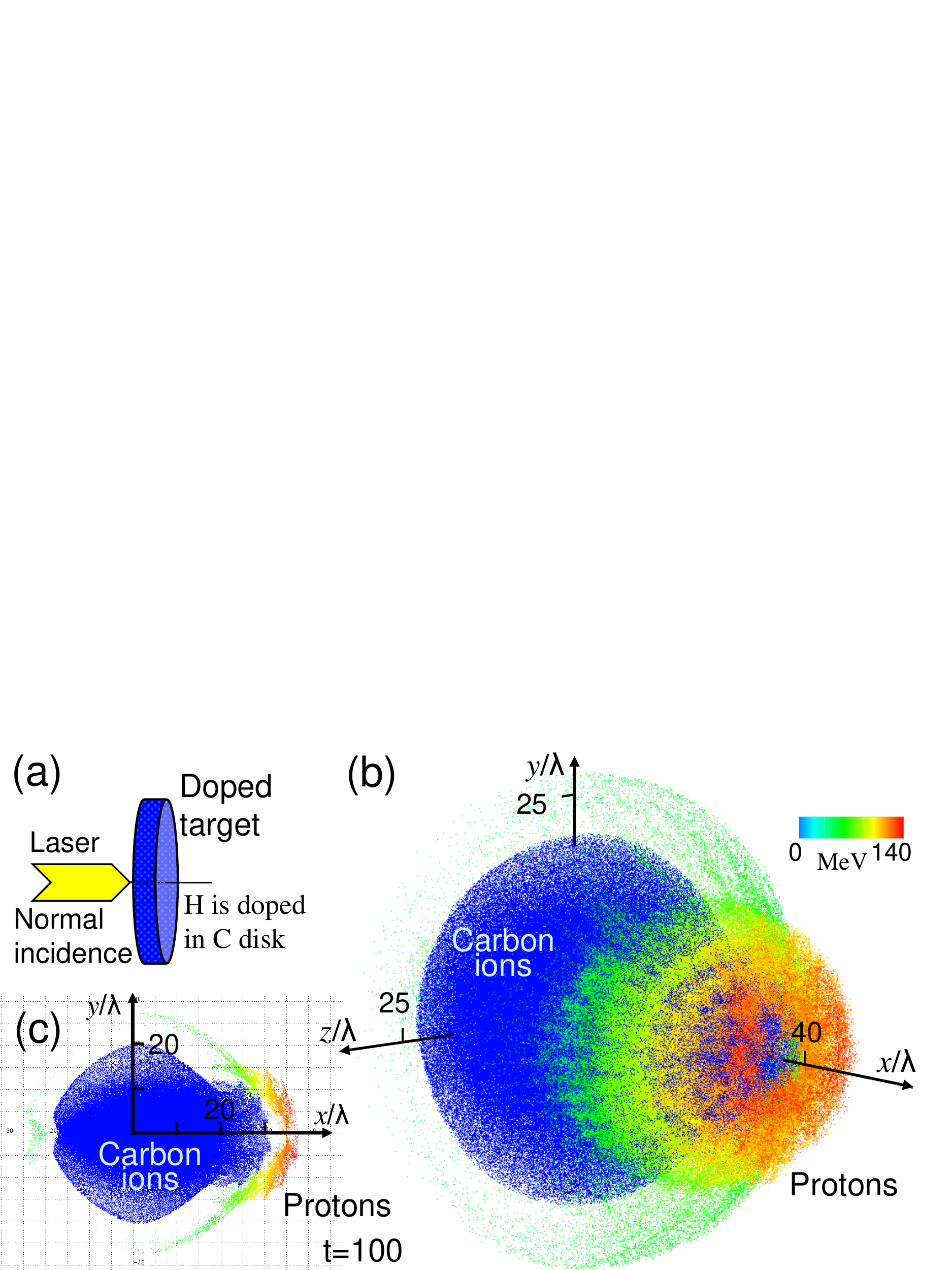}
\caption{
The laser pulse is normally incident on a doped target
that is hydrogen doped in carbon disk (a).
Distribution of the carbon ions and protons (color scale)
at $t=100 T_0$ (b).
The ion cloud, $-2\lambda < z < 2\lambda$
has been taken out to reveal the internal structure;
a two-dimensional projection is shown looking along the $z$ axis (c).
}
\label{fig:fig_b1}
\end{figure}

I show a simulation result which uses a target
that has very low density hydrogen within a disk of high-$Z$ material.
This target can be produced by using a doping technique.
I call this target a doped target in this paper.
Carbon is used for the high-$Z$ material below.
The simulation parameters of this case,
only change the ratio of carbon ions and protons in the target,
the other parameters, target shape and laser pulse conditions,
are the same as the previous case.
It is assumed that the proton density is 1/250 of the previous case.
It is this doped target C$_n$H, $n=170$.
The electron density of the target is the same as previous case,
i.e. $n_{e}=9\times 10^{22}$ cm$^{-3}$.
The proton density is $9\times 10^{19}$ cm$^{-3}$,
the carbon ion density is $1\times 10^{22}$ cm$^{-3}$.
The laser pulse is normally incident on the target
(see Fig. \ref{fig:fig_b1}(a)).
Figure \ref{fig:fig_b1}(b),(c) shows the particle distribution at $t=100 T_0$.
Protons are classified by color in terms of energy.
Figure \ref{fig:fig_b1}(c)
shows a cross section of the ion cloud in the $(x,y)$ plane.
The carbon ions and the protons are clearly separated.
The carbon ion cloud is exploded by Coulomb explosion.
By contrast,
the proton cloud is very narrow.
The protons have moved in the $+x$ direction by RPA
and almost all the protons are positioned at around the $+x$ side surface of
the exploded carbon ions.
The high energy protons gather around the $x$ axis,
which is the laser propagation direction,
and these protons have the same energy level as seen
in Fig. \ref{fig:fig_b1}(c).
The maximum proton energy, $\mathcal{E}_\mathrm{max}$, is 145 MeV.

\begin{figure}[tbp]
\includegraphics[clip,width=11.0cm,bb=5 10 410 188]{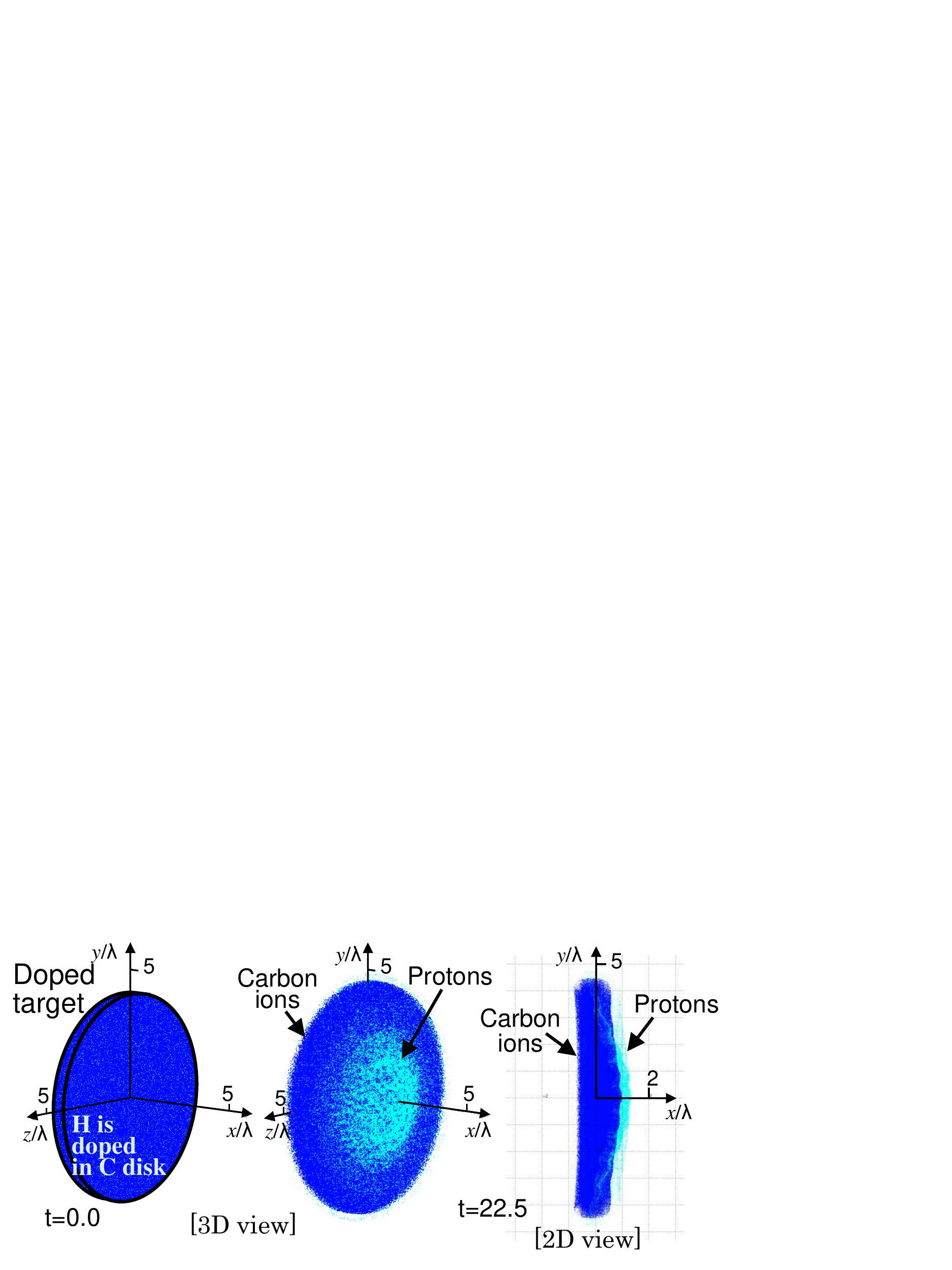}
\caption{
The initial state of the target (t=$0.0$) and
the spatial distribution of ions at an early time ($t=22.5T_0$)
of the doped target.
3D and 2D views are shown.
In the 2D view, half of the box is removed and looking along the $z$ axis.
Dark color shows the carbon ion and
light color shows the proton.
The doped target becomes a double layer target at an early time.
}
\label{fig:fig_b2}
\end{figure}

A proton beam which has a longitudinally narrow distribution in area
and narrow energy spread is obtained.
This is because
the doped target forms a double layer target at an initial stage.
Figure \ref{fig:fig_b2} shows the target state at $t=0$
and early simulation time $t=22.5T_0$ by a 3D view and 2D view.
In the 2D view, half of the box is removed to reveal the internal structure
looking along $z$ axis.
It is shown that the carbon ions and protons are clearly separated
into two layers at $t=22.5T_0$ around the center area of the target.
It forms a double layer target that is coated by a thin and low density proton
layer on the surface of a carbon ion layer.

\begin{figure}[tbp]
\includegraphics[clip,width=8.0cm,bb=10 3 516 398]{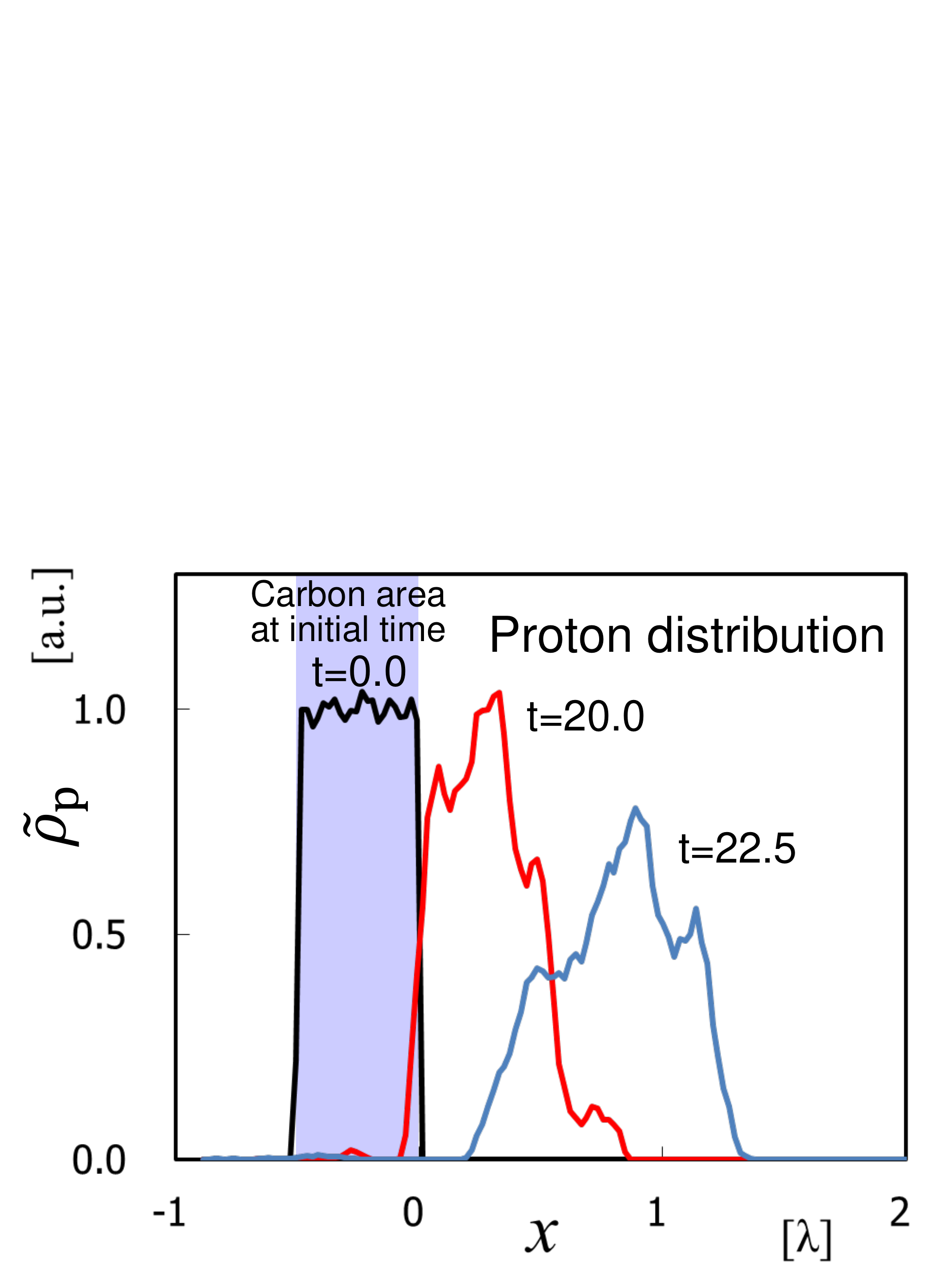}
\caption{
The proton density, $\tilde{\rho}_\mathrm{p}$,
in $x$ coordinate at early simulation time ($t=0.0, 20.0, 22.5T_0$).
It is normalized by the initial proton density, $9\times 10^{19}$ cm$^{-3}$.
Painted area shows the initial position of carbon of the target.
The proton bunch shifts towards the $+x$ direction maintaining its density
and narrow distribution width.
}
\label{fig:fig_b3}
\end{figure}

Figure \ref{fig:fig_b3} shows the proton density along the $x$ coordinate
at each time ($t=0, 20.0, 22.5T_0$).
The vertical axis is the proton density, $\tilde{\rho}_\mathrm{p}$,
which is normalized by its value at $t=0$.
In calculating $\tilde{\rho}_\mathrm{p}$,
an area is selected near the $x$ axis that
has a length from the $x$ axis, $r$, less than $2 \lambda$.
The painted region shows the initial target area,
i.e. the distribution of carbon and hydrogen at the initial time.
The proton bunch moves in the $+x$ direction, laser propagation direction,
maintaining its density and narrow distribution width.
The carbon bunch almost stays at its initial position and shape at this time
(see Fig. \ref{fig:fig_b2}).
Then it forms the double layer target.

\begin{figure}[tbp]
\includegraphics[clip,width=10.0cm,bb=0 3 531 290]{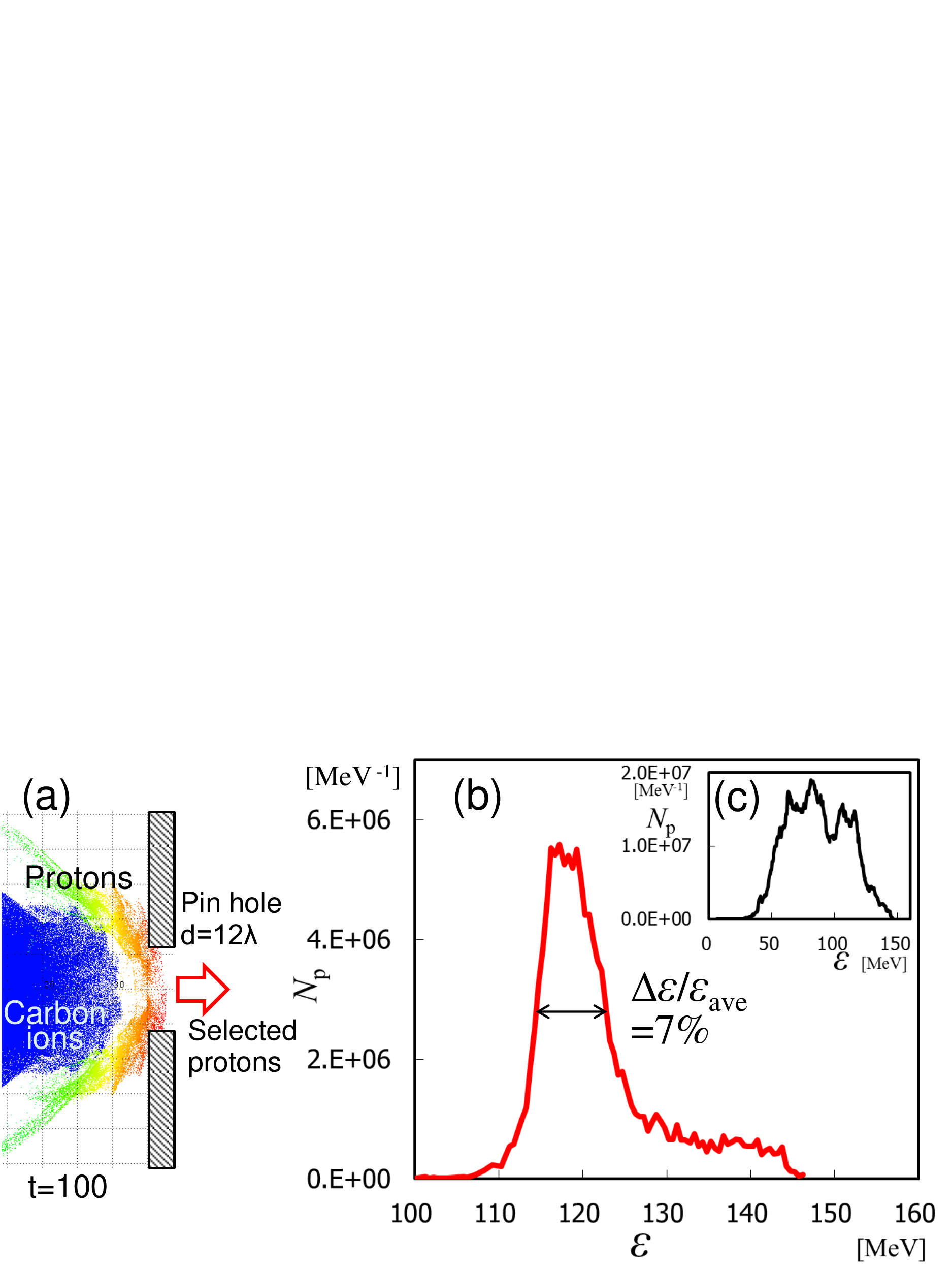}
\caption{
The quasi-monoenergetic proton beam
can be obtained by using a pin hole that is placed front of the target (a).
Energy spectrum of proton beam obtained by using a pin hole (b)
and without the pin hole (c)
at $t=100T_0$.
}
\label{fig:fig_b4}
\end{figure}

Since the high energy protons are distributed in the area around the $x$ axis,
we can obtain only high energy protons by using a pinhole which is placed
on the $x$ axis (see Fig. \ref{fig:fig_b4}(a)).
The diameter of the pinhole is $12 \lambda$.
Figure \ref{fig:fig_b4}(b) shows the energy spectrum of the selected protons
at $t=100T_0$.
We obtain a proton beam with
an average energy of $\mathcal{E}_\mathrm{ave}=120$ MeV with
an energy spread of $\Delta\mathcal{E}/\mathcal{E}_\mathrm{ave}=7\%$
and the proton number is $6\times10^6$ MeV$^{-1}$.

\begin{figure}[tbp]
\includegraphics[clip,width=10.0cm,bb=2 2 533 326]{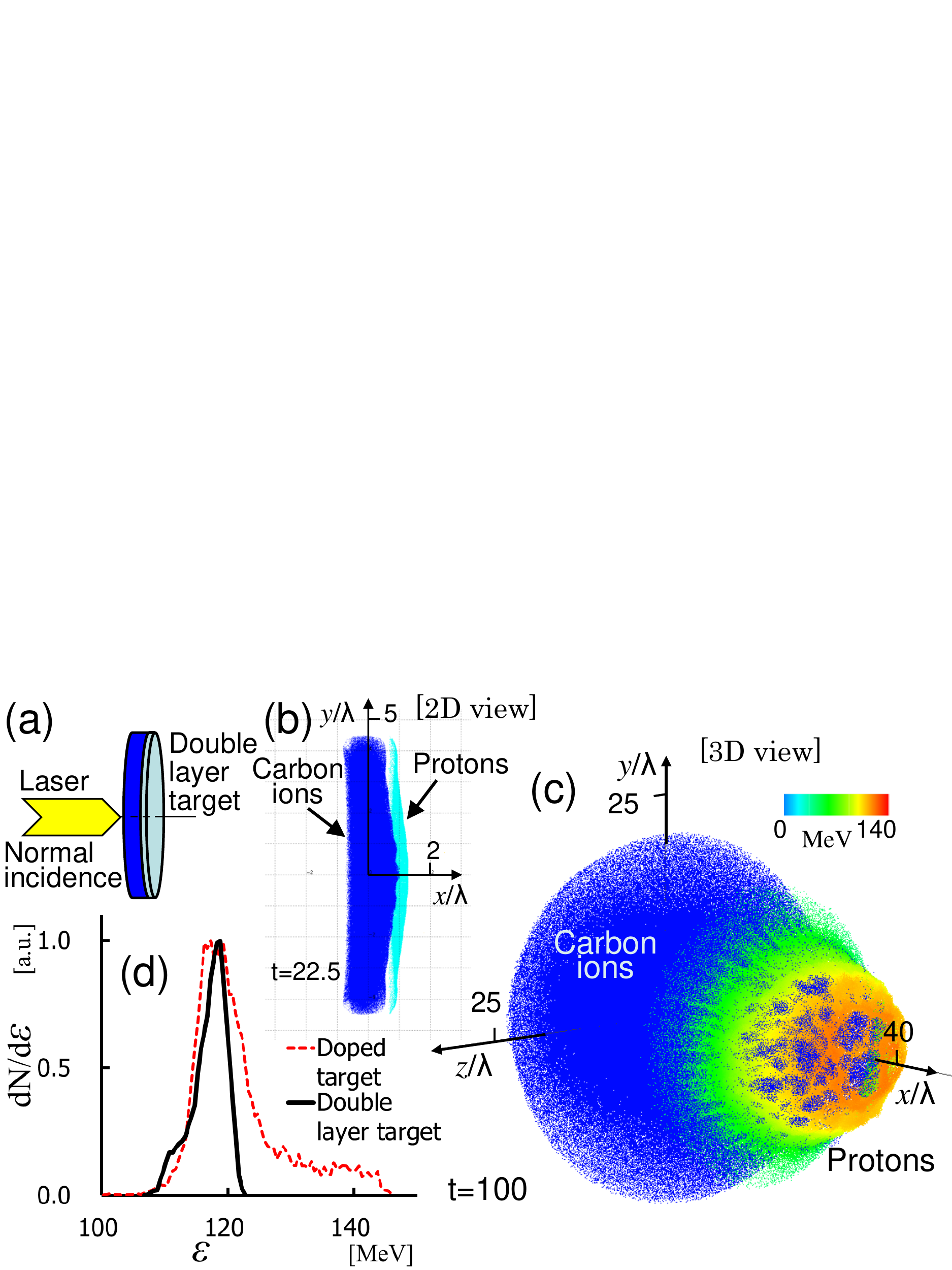}
\caption{
(a) The laser pulse is normally incident on a double layer target.
(b) The spatial distribution of ions at an early simulation time ($t=22.5T_0)$,
shown by 2D view where half of the box is removed
and looking along the $z$ axis.
Dark color indicates the carbon ions and light color the protons.
(c) Distribution of the carbon ions and protons (color scale) at $t=100 T_0$.
For protons, the color corresponds to energy.
(d) Energy spectrum of the proton beam obtained by using a pin hole,
the diameter is $12 \lambda$, at $t=100T_0$.
It is plotted including the previous doped target case for comparison. 
}
\label{fig:fig_b5}
\end{figure}

Next,
I show another simulation result for comparison
with the previous, doped target, case.
It is a double layer target from the beginning.
The simulation parameters are the same as the doped target
without the double layer target. 
The first layer, carbon layer, has the same shape as the previous case target.
The second layer, hydrogen layer is thinner; its thickness is 0.05 $\lambda$,
and the diameter is the same as the first layer
(see Fig. \ref{fig:fig_b5}(a)).
The number of electrons, carbon ions, and protons
is same as the doped target case,
and all the hydrogen of the doped target are moved into the second layer.
That is, the electron density inside the first layer is
$n_{e}=9\times 10^{22}$ cm$^{-3}$,
and carbon ion density is $1\times 10^{22}$ cm$^{-3}$.
Inside the second layer,
the electron density and proton density is $n_{e}=9\times 10^{20}$ cm$^{-3}$.
The other parameters are same as the doped target case.
Figure \ref{fig:fig_b5}(b) shows the target state at $t=22.5T_0$ by a 2D view
where half of the box is removed to reveal the internal structure
looking along $z$ axis.
It is similar with the doped target case (see Fig. \ref{fig:fig_b2}(2D view)).
Figure \ref{fig:fig_b5}(c) shows the particle distribution at $t=100 T_0$.
Protons are classified by color in terms of energy.
The carbon ion cloud is exploded by Coulomb explosion,
although the proton cloud is narrow.
The high energy protons are concentrated around the $x$ axis,
and these protons have the same energy level.
Figure \ref{fig:fig_b5}(d) shows the energy spectrum of the selected protons
by the pinhole which is same as previous case at $t=100T_0$.
It also shows the energy spectrum of the previous doped target case.
Both cases have the same peak point at $\mathcal{E} \approx 120$ MeV,
and narrow energy spread too.
The doped target generates some higher energy, $\mathcal{E} > 130$ MeV,
protons than the double layer target.
It is caused by
the protons being positioned inside the high-$Z$ material layer.
When the protons reach the rear surface of the carbon disk by RPA,
some significant time may have passed compared to other protons.
The electric field of the charged carbon disk is grows in time, and
these protons experience the higher electric field of the charged carbon disk,
and are accelerated to higher energy.

The protons were accelerated in the $+x$ direction in the above consideration.
Here,
I discuss the accelerated direction of a proton within the disk target.
It is considered to be a relatively thick disk, i.e. near cylindrical.
At first, I describe the electric field of the charged cylinder.
I assume that the $x$ axis is in the cylinder height direction,
and the origin is at the center of the cylinder whose
radius is $R$ and height is $\ell$.
The force on charge $q_1$ at the point $x$
on the $x$ axis from a small volume, $dV=rd\theta drdx'$,
at the point $(x',r,\theta)$ of the cylinder is
\begin{equation}
f_{\Delta V}=\frac{1}{4\pi\epsilon_0}
\frac{q_1 q_{_{\Delta V}}}{(x-x')^2+r^2}
\label{fdv}
\end{equation}
where $q_{_{\Delta V}}$ is the charge of the small volume,
$q_{_{\Delta V}}=\rho dV$, $\rho$ is the charge density.
The $x$ component of $f_{\Delta V}$,
$f_{x,\Delta V}=f_{\Delta V} \cdot (x-x')/\sqrt{(x-x')^2+r^2}$.
The total force from the charged cylinder is obtained by
integrating $f_{x,\Delta V}$ over all the cylinder volume $V$.
\begin{equation}
f_x(x) = \int_V f_{x,\Delta V} = \frac{q_1 \rho}{4\pi\epsilon_0}
\int_{-\frac{\ell}{2}}^{\frac{\ell}{2}} \int_0^R \int_0^{2\pi}
\frac{r(x-x')}{\{(x-x')^2+r^2\}^\frac{3}{2}}
d\theta drdx'
\label{fx}
\end{equation}
The $x$ component of the electric  field, $E_x(x)$, is obtained by performing
this multiple integration and dividing by $q_1$.
\begin{eqnarray}
E_x(x)=
\frac{\rho\ell}{2\epsilon_0}
\Biggl[
\sqrt{\Bigl(\frac{x}{\ell}-\frac{1}{2}\Bigr)^2+\Bigl(\frac{R}{\ell}\Bigr)^2}
-\sqrt{\Bigl(\frac{x}{\ell}+\frac{1}{2}\Bigr)^2+\Bigl(\frac{R}{\ell}\Bigr)^2}
+ \left \{ \begin{array}{ll}
-1 & (x < -\frac{\ell}{2}) \\
\frac{2x}{\ell} & ( -\frac{\ell}{2} \leq x \leq \frac{\ell}{2} ) \\
1 & (\frac{\ell}{2} < x)  \\
\end{array}
\right.
\Biggr].
\label{exa}
\end{eqnarray}
The function $E_x(x)$ pass through the origin, $E_x(0)=0$,
and symmetric with respect to the origin, $E_x(-x)=-E_x(x)$.
It is a steadily decreasing function in $x>\frac{\ell}{2}$
and $x<-\frac{\ell}{2}$, and steadily increasing function
in $-\frac{\ell}{2}\leq x \leq \frac{\ell}{2}$.
The formula (\ref{exa}) is written as
\begin{equation}
E_x(x)=E_0\left\{a(x)+b(x)\right\},
\label{exab}
\end{equation}
where $E_0=\rho\ell/2\epsilon_0$,
$a(x)=\sqrt{(x/\ell-1/2)^2+(R/\ell)^2}-\sqrt{(x/\ell+1/2)^2+(R/\ell)^2}$,
$b(x)=\{-1$ for $x<-\ell/2$,
$2x/\ell$ for $-\ell/2 \leq x \leq \ell/2$,
$1$ for $x>\ell/2$.
The function $a(x)$ passes through the origin, $a(0)=0$,
and is symmetric with respect to the origin.
For $\displaystyle \lim_{x \to -\infty} a(x)=1$,
$\displaystyle \lim_{x \to \infty} a(x)=-1$, and decreases monotonically.

The direction of proton acceleration of the disk target
by using this formula, $E_x(x)$, is considered.
Here, I use the coordinate where the origin is located at the center of the
front surface of the disk target.
Therefore, the $x$ component of the electric field at point $x$ is
$E_x(x)=E_0\{a(x-\ell/2)+b(x-\ell/2)\}$, and the conditions of the $x$ value
of the function $b(x)$
change to $(x<0), (0<x<\ell), (\ell<x)$, respectively.
I assumed that the electron layer moves in the $+x$ direction by $\Delta x$
maintaining the initial shape
and that the ion layer stays at the initial position with its initial shape.
We can obtain the electric field at the $x$ position in this situation
by adding the electric field of the ion disk and the electron disk.

\begin{figure}[tbp]
\includegraphics[clip,width=8.0cm,bb=5 2 373 300]{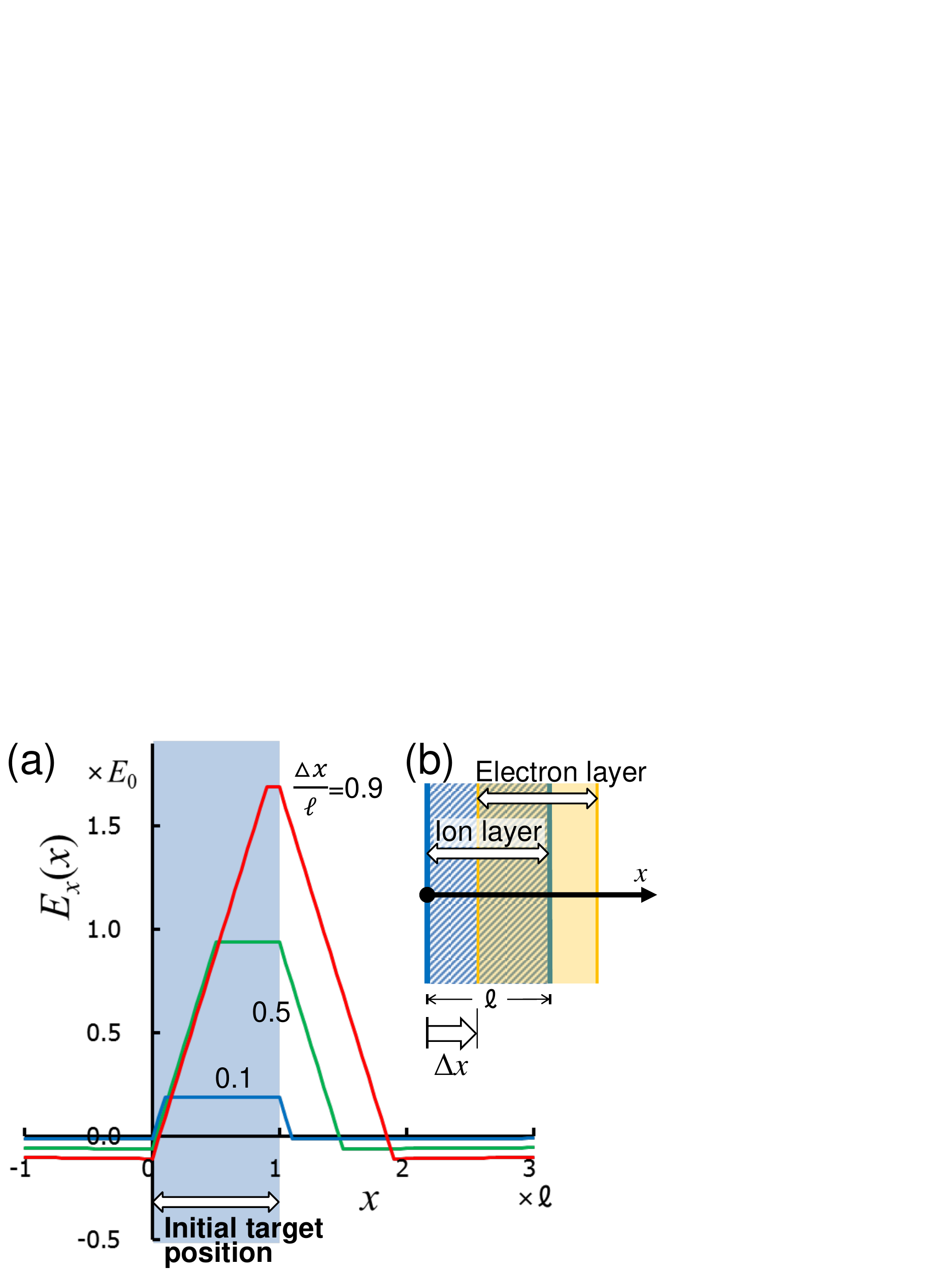}
\caption{
The electric field in the $x$ direction, $E_x(x)$, in the target and
near the target (a),
when the electron layer moves in the $x$ direction by $\Delta x$ (b).
The electric fields are shown by the amount of the movement of
the electron layer $\Delta x/\ell = 0.1,0.5,0.9$.
$E_x(x)$ in the target
has almost all positive values in all cases,
which means that protons in the target are accelerated forward in the
laser propagation, $+x$, direction.
}
\label{fig:fig_b6}
\end{figure}

When the ion layer and the electron layer overlap, $0<\Delta x<\ell$,
the electric field, $E_x(x)$, is
\begin{eqnarray}
E_x(x)=E_0 \Biggl[ a(x-\frac{\ell}{2})-a(x-\frac{\ell}{2}-\Delta x) +
\left \{ \begin{array}{ll}
0 & (x<0, \: \ell+\Delta x<x) \\
\frac{2x}{\ell} & (0<x<\Delta x) \\
\frac{2\Delta x}{\ell} & (\Delta x<x<\ell)  \\
2-\frac{2(x-\Delta x)}{\ell} & (\ell<x<\ell+\Delta x)
\end{array}
\right.
\Biggr].
\label{exlap}
\end{eqnarray}
This function is shown in Fig. \ref{fig:fig_b6}(a) which in the cases of
the amount of the electron layer shift is small, medium, and large,
which are $\Delta x/\ell=$ 0.1, 0.5, and 0.9 respectively.
I take $R/\ell=4\lambda/0.5\lambda=8$
which is the value of the previous simulations.
The electric field, $E_x(x)$, is almost all positive in the ion layer,
which means that protons in the target are accelerated in the $+x$,
laser propagation, direction.
The electric field in the ion layer grows with $\Delta x$.
On the other hand for $x<0$ the electric field is negative,
therefore the protons where $x<0$ are accelerated in the $-x$ direction.

When the electron layer leaves the ion layer, $\Delta x > \ell$,
in statements formula (\ref{exlap}) $\frac{2\Delta x}{\ell}$ changes to 2,
and the conditional change so that
$\Delta x \rightarrow \ell$, $\ell \rightarrow \Delta x$.

It was assumed that $\Delta x >0$, since the electron layer moves in
the $+x$ direction.
Therefore, $E_x(x)<0$ where $x<0$ for any $\Delta x$.
This results from formula (\ref{exlap}) in the case ($x<0$)
and the formula of $\Delta x>\ell$ case.
This is because $a(x)$ is a steadily decreasing function.
And, $E_x(\Delta x/2)\ge 0$ for any $\Delta x$.
Therefore, the point where $E_x(x)=0$ exists in $0<x<\Delta x/2$.
The point $x$ is obtained by solving
$E_x(x)=E_0\{a(x-\ell/2)-a(x-\ell/2-\Delta x)+2x/\ell\}=0$.
Assuming $(\ell/2R)^2 \ll 1$, and $x$ is same order
with or smaller than $\ell$,
we obtain $a(x) \approx -x/R$.
Therefore, $a(x-\ell/2) \approx \ell/2R -x/R$
and $a(x-\ell/2-\Delta x) \approx \ell/2R-x/R+\Delta x/R$,
I assumed $\Delta x$ is the same order with or smaller than $\ell$.
By using this, we obtain the $x$ value that the point $E_x(x)=0$,
\begin{equation}
x_0=\frac{\ell}{2R} \Delta x
\label{ex0}
\end{equation}
$x_0=0$ at $t=0$ and grows in time,
because $\Delta x=0$ at $t=0$ and it grows in time. 
The disk target which is used in our simulations is $\ell/2R = 0.06$.
Therefore, $x_0 \ll \Delta x$.
The point $x_0$ is located almost at the front surface of the target,
$x_0 \approx 0$, at early times.
The point $x_0$ move in the $+x$ direction in time
and has the speed $v_0=x_0/\Delta t$.
Since, the electron layer speed $v_e=\Delta x/\Delta t$,
then $v_0/v_e=\ell/2R =0.06$ in our simulation cases.
This means, the velocity of the $x_0$ point is very slow
compared with the electron layer speed.
Even the protons which are located at $x \approx 0$,
near the front surface of the target,
if the protons get a higher speed than $v_0$ by RPA,
the protons are accelerated in the $+x$ direction.
This condition, $v_p(t)>v_0(t)$,
becomes less strict with distance of the protons from $x=0$,
where  $v_p$ is the speed of the protons.
In the radiation pressure dominant acceleration (RPDA),
the protons obtain a relatively high momentum, velocity, at initial times.
By the above considerations,
the protons where $x>0$ are accelerated in the $+x$ direction and
the protons where $x<0$ are accelerated in the $-x$ direction.
This consideration is suited to that where
radiation pressure acceleration is dominant at an early acceleration stage.

The acceleration direction of a proton changes if its position is
$x>0$ or $x<0$.
Here, I consider the case of the hydrogen layer is located around $x=0$. 
It is the hydrogen layer on the front side surface of a high-$Z$ atom layer.
When the electron density of the hydrogen layer is similar to the
a high-$Z$ atom layer,
the origin in the above consideration
is considered to be the front side surface of the hydrogen layer.
That means, the protons are positioned in $x>0$ region.
Therefore, protons are accelerated forward in the laser propagation direction.
When the electron density of the hydrogen layer is negligibly small
compared with the high-$Z$ atom layer, the origin is considered
to be the front side surface of the high-$Z$ atom layer.
Since, the protons are positioned in $x<0$ region, protons are accelerated
in the opposite direction with respect to the laser propagation. \cite{TM}

I use the disk target in the above consideration of the accelerated direction
of the proton.
However, this not only for the disk target,
but for the foil target.
This is because we can consider the disk to be located at the position of
the laser spot on the foil.

\section{CONCLUSIONS}

Proton acceleration driven by a laser pulse irradiating a disk target,
a CH$_2$ target and a doped target,
is investigated with the help of 3D PIC simulations.
In the initial acceleration stage, RPA plays an important role
which forms a two layer target, double layer target.
Even for a laser intensity of $5\times10^{21}$ W/cm$^2$
which is not enough for RPDA.
Other acceleration schemes after it,
the Coulomb explosion of each ion cloud and
Coulomb repulsion between each cloud,
generate a high energy and quasi-monoenergetic proton beam.

In a polyethylene (CH$_2$) target,
a proton layer and a carbon ion layer are formed.
In this separated situation, a strong Coulomb explosion of each layer
and Coulomb repulsion between each layer generate the high energy protons.
We can obtain a 210MeV proton beam in the oblique incidence case.
The doped target
form a double layer target
and it generates a quasi-monoenergetic proton beam.
There may be a difficulty to make the target which has a thin and low density
hydrogen layer on some ion layer,
but it is made automatically by RPA.
It is not necessary to prepare the double layer target.
In this paper,
I used carbon for the high-$Z$ material of the doped target,
we can get this type of target by using Diamond-like carbon (DLC)
which includes low density hydrogen in a carbon foil.
The thickness of the target in our simulations corresponded to 0.1$\mu$m
with ordinary solid polyethylene and carbon
which has the same value of $\int n dx$, where $n$ is the mass density.
We can generate over 200MeV protons by using a 620TW,
18J laser pulse of peak intensity $5\times10^{21}$ W/cm$^2$.

\section*{ACKNOWLEDGMENTS}
I thank P. Bolton, S. V. Bulanov, T. Esirkepov, M. Kando, J. Koga, K. Kondo,
and M. Yamagiwa for useful discussions.
The computations were performed using the PRIMERGY BX900 supercomputer
at JAEA Tokai.
This work was supported by JSPS KAKENHI Grant Number 23540584.


\end{document}